\documentclass[preprint,showpacs,preprintnumbers,amsmath,amssymb,prl,aps]{revtex4-1}

\usepackage{epsfig}
\usepackage{graphicx}
\usepackage{bm}
\usepackage{float} 

\begin{document}

\title{Aging of the surface of Bi$_2$Se$_3$}

\author{Deepnarayan Biswas$^1$}
\author{Sangeeta Thakur$^1$}
\author{Khadiza Ali$^1$}
\author{Geetha Balakrishnan$^2$}
\author{Kalobaran Maiti$^{1}$\footnote{Corresponding author: kbmaiti@tifr.res.in}}

\affiliation{$^1$Department of Condensed Matter Physics and
Materials' Science, Tata Institute of Fundamental Research, Homi
Bhabha Road, Colaba, Mumbai - 400 005, India.\\
$^2$Department of Physics, University of Warwick, Coventry, CV4 7AL,
UK.}

\date{\today}

\begin{abstract}
We study the electronic structure of Bi$_2$Se$_3$ employing high
resolution photoemission spectroscopy and discover the dependence of
the behavior of Dirac particles on surface terminations. The Dirac
cone apex appears at different energies and exhibits contrasting
shift on Bi and Se terminated surface with complex time dependence
emerging from subtle adsorbed oxygen-surface atom interactions.
These results uncover the surface states behavior of real systems
possessing topologically ordered surface.
\end{abstract}



\maketitle

Topological insulators are like ordinary insulators in the bulk with
gapless surface states protected by time reversal symmetry
\cite{hasan_rev,3dti}. These materials have drawn much attention in
the recent times followed by the proposals of the realization of
exotic physics involving Majorana Fermions \cite{majorana}, magnetic
monopoles \cite{monopole} etc. In addition to such fundamental
interests, the predicted special properties of these states make
them useful for the technological applications ranging from
spintronics to quantum computations. Although the topological states
are protected by time reversal symmetry\cite{aging_hasan}, numerous
experiments show instability of the topological states with aging.
Plethora of contrasting scenarios, anomaly on absorption of foreign
elements, etc. are observed in the experimental
studies\cite{aging_hasan,aging_kong}. It is evident that the real
materials are complex and may not be commensurate to the theoretical
predictions that makes this issue an outstanding problem in various
branches of science and technology.

In order to elucidate these puzzles in real materials, we studied
the electronic structure of a typical topological insulator,
Bi$_2$Se$_3$ at different experimental conditions such as the
behavior of differently terminated surface, evolution of the
electronic structure on aging, etc. employing high resolution
photoemission spectroscopy. Bi$_2$Se$_3$ forms in a layered
structure (see Fig. 1(a)) with the quintuple layers of
Se-Bi-Se-Bi-Se stacked together by Van der Waals force
\cite{surface_xrd}. The surface electronic structure exhibits
topological order with the apex of the Dirac cone, called Dirac
Point (DP) at finite binding energies due to finite charge carrier
doping arising from impurities, imperfections, etc.
\cite{contradict_rader,stable_atuchin,stable_goly} These states
often show instability with time and complex time evolutions
\cite{aging_hasan,aging_kong}, which has been attributed to
different phenomena such as relaxation of Van der Waals bond
\cite{relaxation_noh}, the surface band bending \cite{aging_hasan},
dangling surface states \cite{tanmoy}, etc. There exists contrasting
arguments indicating the necessity of unusually large change in the
bond length for the relaxation of Van der Waals bonds
\cite{dft_fukai}. Evidently, the real materials exhibit significant
deviations from theoretical wisdom
\cite{BB_bianchi,dft_wang,BB_zhang} albeit the electronic states
being topologically protected. Here, we discover that the behavior
of the topological states is dependent on surface terminations. The
anomalies in the behavior of the Dirac particles actually depends on
subtle interactions of the adsorbates with the surface atoms.

\begin{figure}
 \vspace{-2ex}
\includegraphics [scale=0.6]{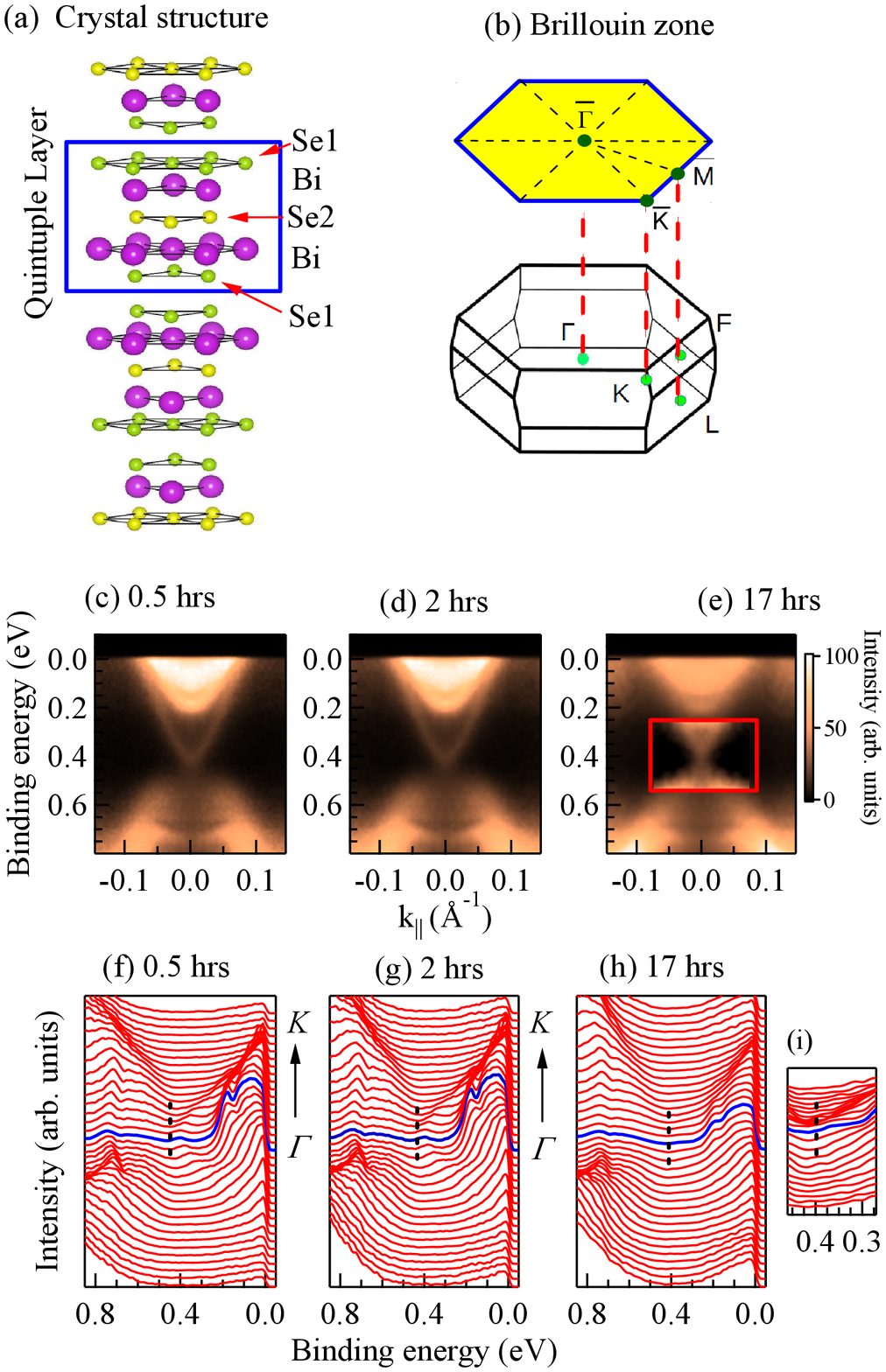}
\vspace{-4ex}
 \caption{(a) Crystal structure of Bi$_2$Se$_3$ exhibiting stack of
quintuple layers of Bi and Se and corresponding (b) Brillouin zone.
ARPES data at 20 K along $\Gamma-K$ for 'Clv2' after (c) 0.5 hrs,
(d) 2 hrs and (e) 17 hours of cleaving. (f) - (g) show the
corresponding energy density curves. (i) Rescaled spectral region
exhibiting the Dirac point with better clarity.}
\end{figure}

In Fig. 1, we show the signature of Dirac cone representing
the topological surface states in the angle resolved photoemission
spectroscopic (ARPES) data. The high symmetry points and the
Brillouin zone defined in the reciprocal space of Bi$_2$Se$_3$ are
shown in Fig. 1(b). In addition to the metallic Dirac states,
several bulk bands cross the Fermi level, $\epsilon_F$
indicating metallicity of the bulk electronic structure
\cite{hasan_rev,3dti}. Curiously, the Dirac point appears at an
unusually high binding energy of about 0.45 eV. To ascertain the
reproducibility of these data, the sample was cleaved several times
and measured in identical conditions. We discover results of two
categories. (i) In one case, DP appears around 0.3 eV binding energy
as shown in Fig. 2, consistent with the earlier results
\cite{hasan_rev,3dti,aging_hasan,aging_kong}. We denote this case as
'Clv1'. (ii) In the other case denoted as 'Clv2', the DP appears
around 0.45 eV binding energy. Time evolution of the 'Clv2' spectra
at 20 K is shown in Figs. 1(c) - 1(e) and corresponding energy
distribution curves (EDC) in 1(f) - 1(h). Ironically, the DP shifts
towards $\epsilon_F$ with the increase in time delay from cleaving
suggesting an effective hole doping with time and/or passivation of
the electron doped bulk with aging.

\begin{figure}
 \vspace{-2ex}
\includegraphics [scale=0.6]{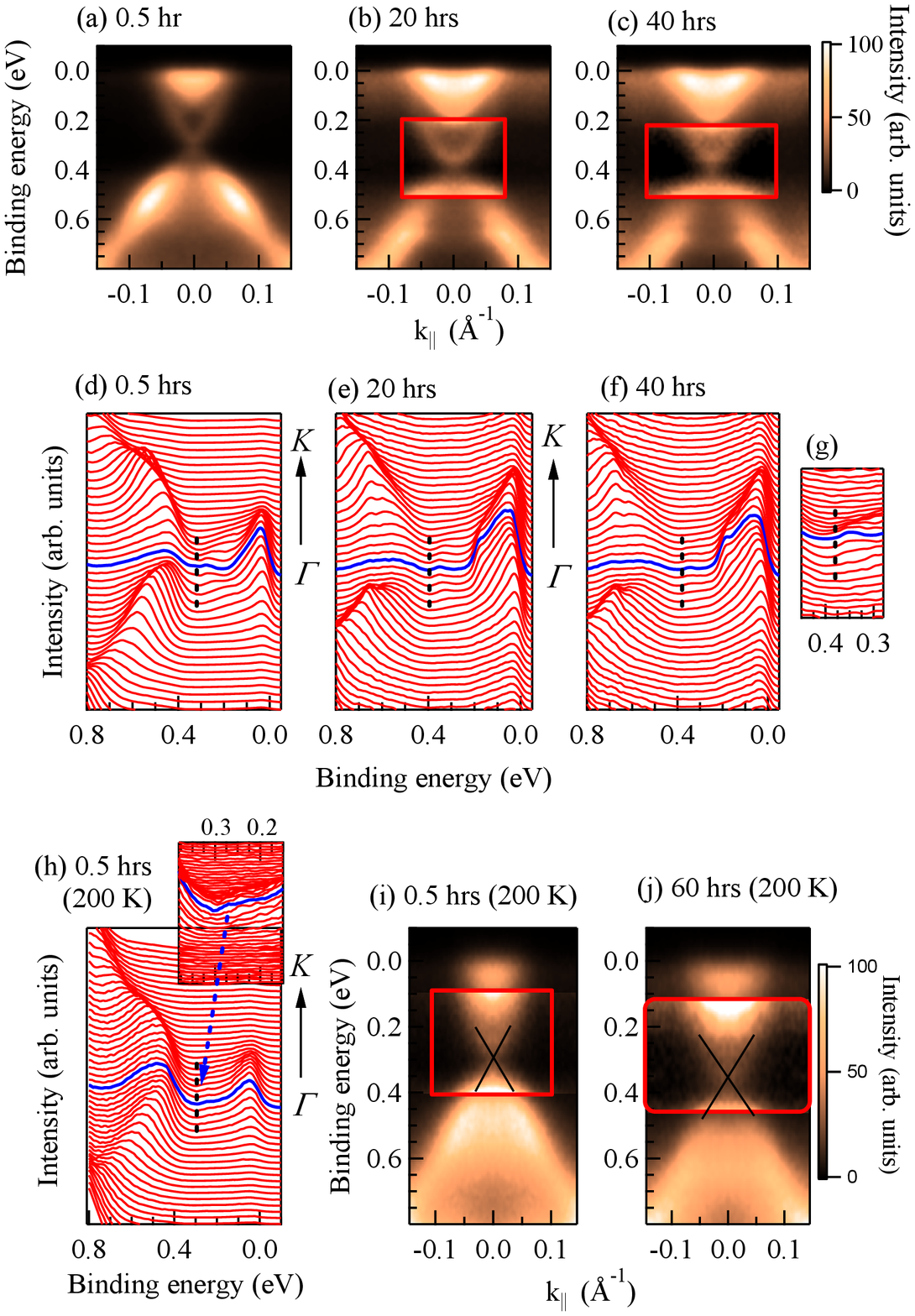}
\vspace{-4ex}
 \caption{(a) ARPES data at 20 K along $\Gamma-K$ direction from Clv1 surface
after (a) 0.5 hrs, (b) 20 hrs and (c) 40 hours after cleaving. (d) -
(f) show the corresponding energy density curves. (g) Rescaled
spectral region near Dirac point. (h) EDC at 200 K. ARPES data at
200 K at (i) 0.5 hrs and (j) 60 hours after cleaving.}
\end{figure}

DP in the 'Clv1' spectra shown in Fig. 2 appears around 0.3 eV for
freshly cleaved surface and gradually shifts away from $\epsilon_F$
with the increase in time delay suggesting an electron
doping with time. Evidently, the contrasting scenario for Clv1 and
Clv2 surface is curious. DP appears to stabilize at a long time
delay in both the cases. The experiments at 200 K exhibit similar
trend in energy shift with a slightly different saturation value.
All these results indicate that the cleaved surfaces are
qualitatively different in the two cases with $\epsilon_F$ pinned at
different energies and anomalous shift of the Dirac point with time.
Some element specific study is necessary to reveal the surface
chemistry of these materials.

\begin{figure}
 \vspace{-2ex}
\includegraphics [scale=0.6]{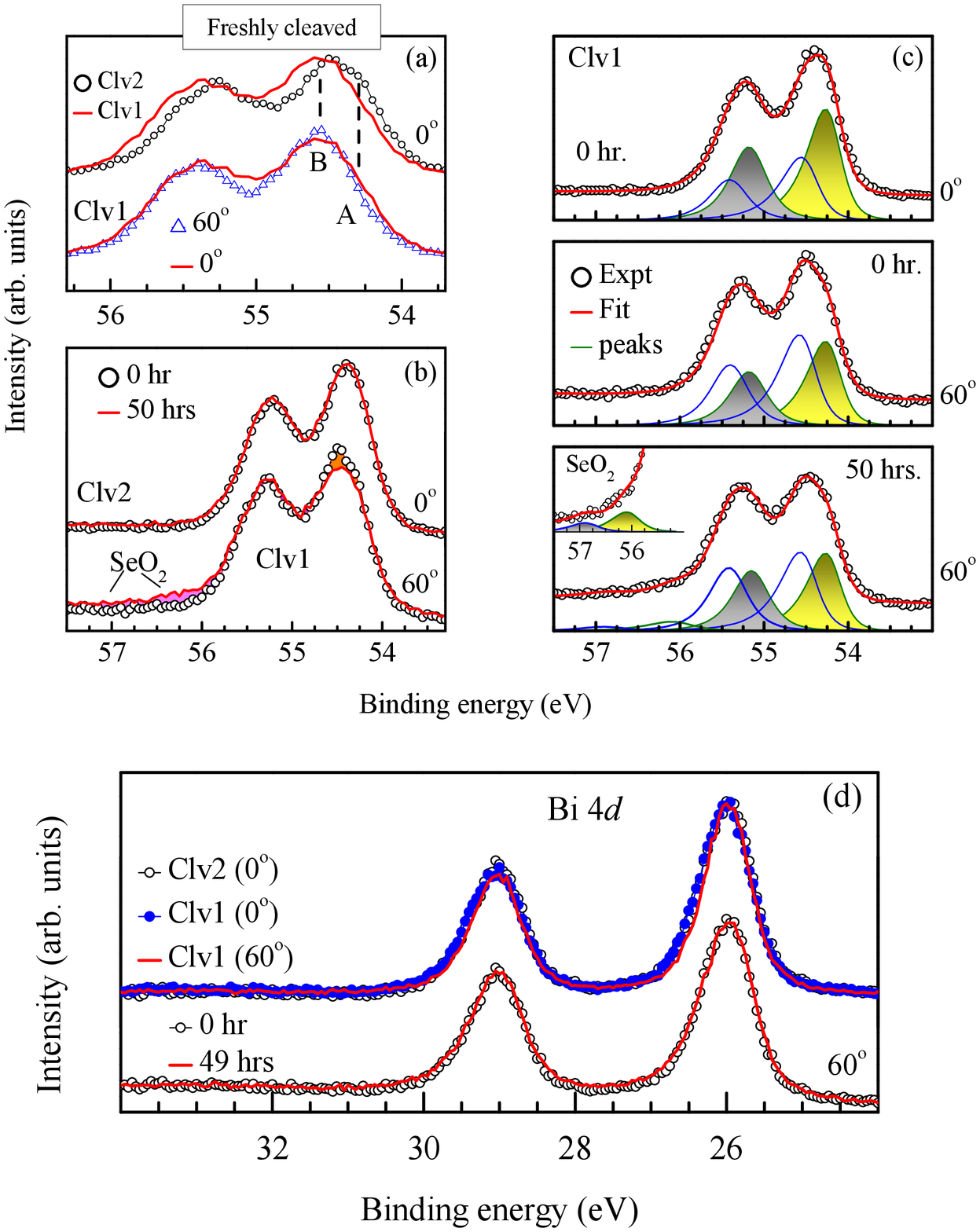}
\vspace{-2ex}
 \caption{(a) Se 3$d$ spectra from Clv1 (line) and Clv2 (open circles)
surfaces at normal emission. The open triangles are the Clv1 data at
60$^o$ off-normal electron emission. (b) Se 3$d$ spectra at normal
emission from Clv2 surface and at 60$^o$ off-normal emission from
'Clv1' surface. The data from freshly cleaved surface and after 50
hours of delay are shown by lines and open circles, respectively.
(c) The fit of the Se 3$d$ spectra from Clv1 at different emission
angles and time delay. (d) Bi 4$d$ spectra from Clv1 and Clv2
surfaces at different emission angles and delay times exhibiting
identical lineshape for every case.}
\end{figure}

We employed $x$-ray photoemission (XP) spectroscopy to probe the
surface chemistry \cite{cuprate}. The normal emission Se 3$d$
spectrum from Clv2 surface exhibits two peak structures for each
spin-orbit split peaks as denoted by 'A' (54.3 eV) and 'B' (54.6 eV)
in Fig. 3(a) for the 3$d_{5/2}$ photoemission signal. The feature, A
becomes significantly weaker in Clv1 spectra with subsequent
enhancement of the feature B. The peak B enhances further in the
spectra collected at 60$^o$ off-normal emission from Clv1 surface.
Off-normal emission makes the technique more surface sensitive
\cite{manju,rabi} (the photoemission probing depth, $d = \lambda
cos\theta$, $\lambda$ = escape depth and $\theta$ = emission angle
with surface normal). The enhancement of B with the increase in
surface sensitivity suggests its surface character, thereby,
assigning the feature A to the bulk Se photoemission. Thus, the
cleaving of Bi$_2$Se$_3$ leads to different surface terminations
exposing Se in Clv1 case and Bi in Clv2 case. We note here that top
post removal method was necessary to prepare well ordered clean
sample surface exhibiting significantly strong binding between the
quintuple layers.

The second important observation is shown in Fig. 3(b), where the
60$^o$ angled emission spectra from Clv1 surface exhibit a decrease
of B and subsequent increase of two weak features at higher binding
energies (marked `SeO$_2$' in the figure) indicating emergence of a
new kind of Se at the cost of some of the surface Se species. The
changes in the normal emission spectra from Clv2 surface, however,
are not distinct indicating weak influence of aging on the
subsurface Se species. The features deriving the Se 3$d$ spectra
were simulated by a set of asymmetric peaks as shown in Fig. 3(c).
The shaded peaks represent the bulk Se photoemission and the other
features correspond to the surface Se. The peak position of the
emerging Se components with time and their intensity ratio indicate
their origin to Se 3$d$ signals from SeO$_2$ species\cite{seo2}. Bi
4$d$ spectra shown in Fig. 3(d), however, exhibit quite similar
lineshape at various experimental conditions employed.

\begin{figure}
 \vspace{-2ex}
\includegraphics [scale=0.6]{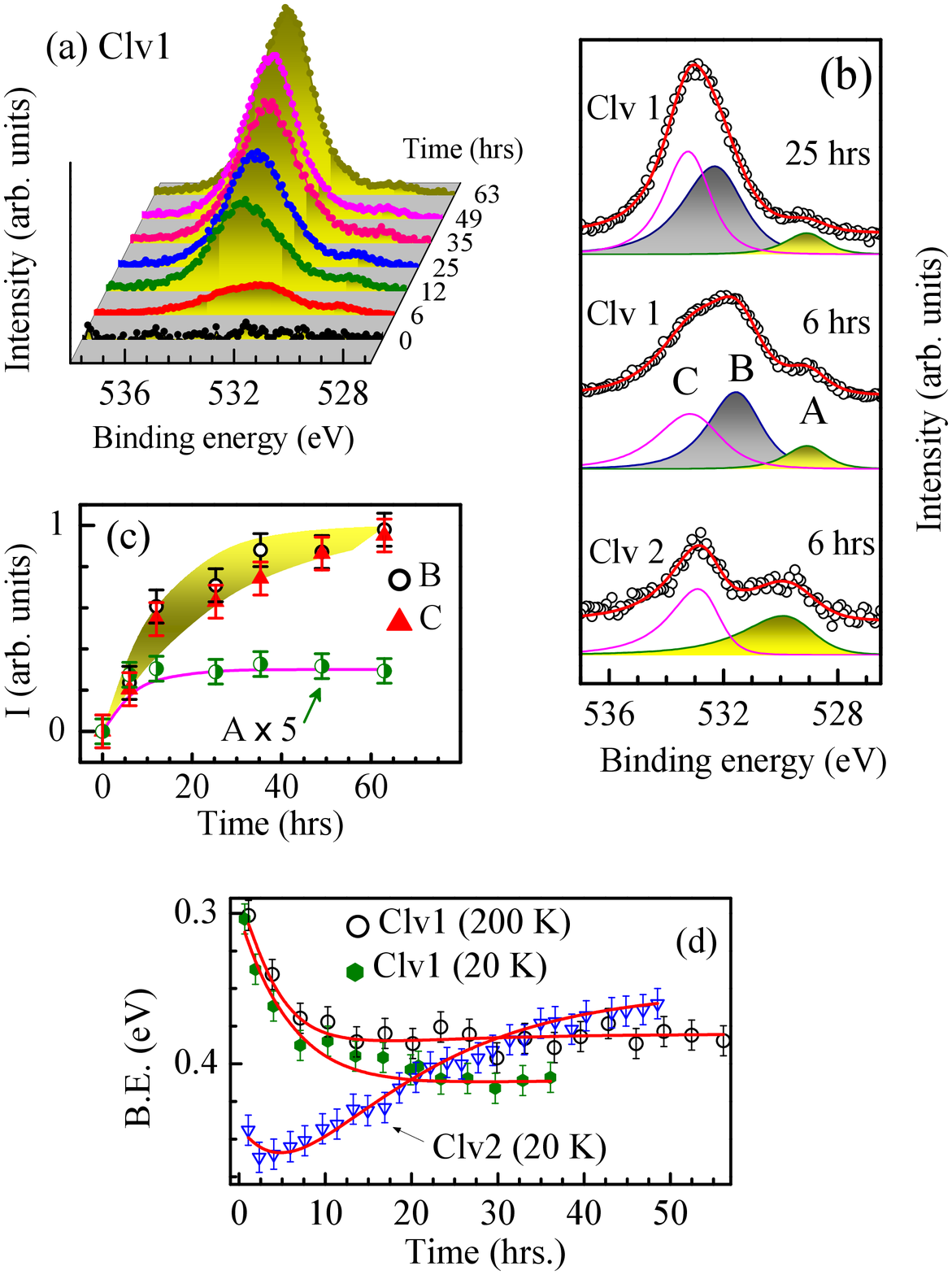}
\vspace{-2ex}
 \caption{(a) Growth of O 1$s$ feature on Clv1 surface with time at 20 K.
(b) Fit of the O 1s spectra from both Clv1 and Clv2 surfaces with
asymmetric peaks. Except the feature A, all the features of 25 hrs
delay spectrum are compressed by 2 times for clarity. (c) Evolution
of the O 1$s$ intensities with time. The line and the shaded region
show exponential time dependence. (d) Binding energy shift of the
Dirac point with time (symbols). The lines show exponential fits.}
\end{figure}

Experiments on freshly cleaved surface do not show signature of
impurity features. The oxygen 1$s$ signal appears to emerge with
aging and gradually grows with the increase in delay time. A
representative case is shown in Fig. 4(a) for Clv1 surface after
normalizing by the number of scans. The Clv1 spectrum after 6 hours
delay exhibits three distinct features denoted by A, B and C in Fig.
4(b). The Clv2 spectrum at about 6 hours delay exhibits relatively
more intense A and weaker C with no trace of B suggesting different
characteristics of adsorbed oxygens on different surfaces. In both
the cases, the feature A grows quickly relative to the other
features as shown in Fig. 4(c).

Now, the question is, if this surface modification influences Dirac
states although they are protected by topological order. In Fig.
4(d), we show the evolution of the Dirac point with aging and
observe that the DP in Clv1 spectra appears around 0.3 eV, on
freshly cleaved surface. With increase in time delay, it gradually
shifts towards higher binding energies with time and stabilizes
around 0.4 eV. The same set of experiments at 200 K exhibit DP
around 0.3 eV along with a weaker energy shift saturating around
0.38 eV that opens up new possibilities in understanding the
behavior of these exotic states vis a vis existing wisdoms
\cite{BB_zhang,h2o_benia,co_bianchi}. Ironically, the Clv2 spectra
exhibit a reverse scenario with DP appearing at much higher binding
energy of 0.45 eV and shifting in the opposite direction implying an
effective hole doping case. The time evolution of DP can be
expressed as,

$$\epsilon_{DP}(t) = \epsilon_0 - \alpha \times exp(-{t/t_{k1}}) +
\beta \times exp(-{t/t_{k2}})$$

Here, $\epsilon_0$ = DP at long delay time, $t$. $t_{k1}$ and
$t_{k2}$ are the time constants. $\alpha$ and $\beta$ are positive
and are related to the electron and hole doping, respectively. The
data points at different conditions can be captured remarkably well
with the above equation. Now, at $t = 0$, the binding energy at DP
is $(\epsilon_0 - \alpha + \beta)$. Therefore, the shift of DP can
be expressed as
$$\Delta\epsilon(t) = \alpha [1-\exp(-{t/t_{k1}})] - \beta [1 - exp(-{t/t_{k2}})]$$
values of $\epsilon_0$, $\alpha$ and $\beta$ are 0.41, 0.13 \& 0.02
for 'Clv1' at 20 K; 0.38, 0.12 \& 0.02 for 'Clv1' at 200 K, and
0.35, 0.16 \& 0.25 for 'Clv2' at 20 K. The parameters are quite
similar in all the cases except a large $\beta$ for the 'Clv2' case.
$t_{k1}$ and $t_{k2}$ are 6 hrs and 16 hours at 20 K in both the
cleaved cases. $t_{k1}$ becomes 4 hrs at 200 K leaving $t_{k2}$
unchanged. The growth of oxygen can be expressed as $[1 -
exp(-t/t_k)]$ with $t_k$ = 16 hrs for the features B \& C, and 6 hrs
for A at 20 K indicating a possible link between the DP shift and
oxygen growth.

Since the time delay primarily modifies the surface, the chemical
potential shift must be due to the change in electron count in the
surface electronic structure \cite{cab6}. Three types of oxygen species are
found to grow on the surface. The feature A grows quickly and have
dominant contribution on Clv2 (Bi terminated) surface. The feature B
is absent in Clv2 spectra and have large contribution in Clv1
spectra (Se-terminated surface) suggesting its bonding with surface
Se leading to electron doping, while the feature A corresponds to
oxygen bonded to Bi leading to hole doping.

\begin{figure}
 \vspace{-8ex}
\includegraphics [scale=0.6]{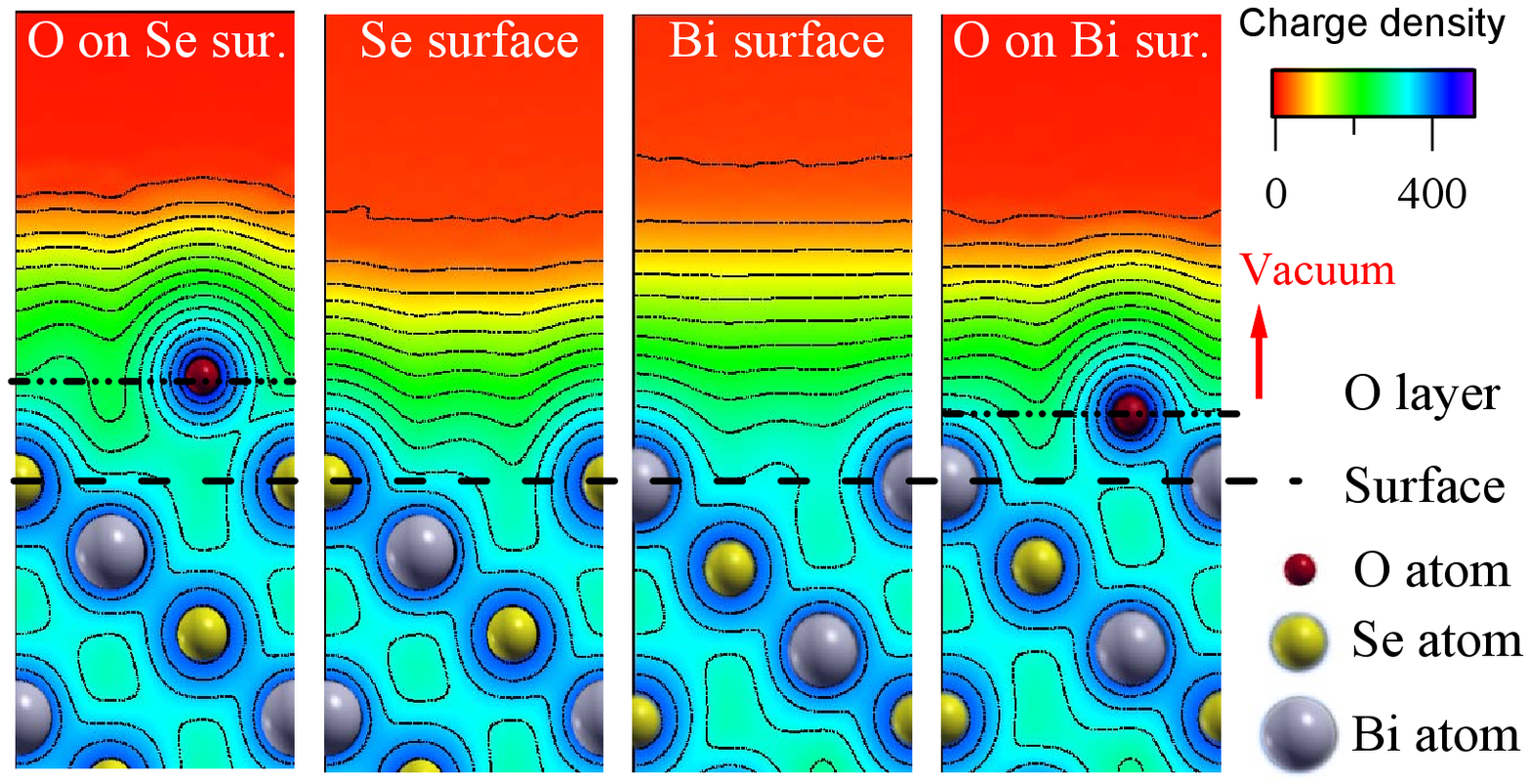}
\vspace{-36ex}
 \caption{Calculated charge density plots (left to right) for
oxygen on Se-terminated surface, Se terminated surface, Bi
terminated surface and oxygen on Bi-terminated surface.}
\end{figure}

The charge densities on the surface are calculated using full
potential density functional theory and shown in Fig. 5. The Bi
terminated surface exhibit highly extended electron density
spreading over a larger spatial distance away from the surface
compared to the Se-terminated case. The Fermi surface corresponding
to the Bi-terminated surface is significantly larger than that
corresponding to Se-terminated case. Thus, DP is expected to appear
at higher binding energy in Bi-terminated case compared to the
Se-terminated case as found experimentally.

Se and O belong to the same group of the Periodic table with O being
the topmost element with higher electronegativity. Therefore, O on
Se surface forms SeO$_x$ complex (signature of SeO$_2$ appears in
the Se 3$d$ spectra). Se-O bonding will attract electron cloud from
the neighborhood reducing the electron density in the Bi-Se
neighborhoods as manifested in the left panel of Fig. 5 by increase
in spatial charge density contours around oxygen sites relative to
the pristine case. Since the conduction band consists of $p$
electrons, this would lead to an effective electron doping in the
conduction band. On the other hand, oxygen on Bi-terminated surface
would form BiO$_x$ complexes leading to more charge localization in
the vicinity of oxygens (see right most panel of Fig. 5) reducing
the Fermi surface volume; an effective hole doping scenario.

In summary, we studied the surface electronic structure of a
topological insulator, Bi$_2$Se$_3$ employing high resolution
photoemission spectroscopy. We observe the sensitivity of the Dirac
states on surface termination. The surface states and the impurities
appear to play a complex role leading to complex Fermi surface
reconstructions emerging as a shift of the Dirac cone. These
materials have been drawing much attention due to their potential
technological applications in addition to the fundamental issues of
realizing magnetic monopoles, the observation of Majorana fermions,
etc. The results presented here reveal the complex microscopic
details of the surface states necessary in the realization of such
ambitious projects in real materials.

\section*{Method}

High quality single crystals of Bi$_2$Se$_3$ were prepared by
Bridgeman method and characterized using $x$-ray Laue diffraction.
Angle resolved photoemission (ARPES) measurements were carried out
employing Gammadata Scienta R4000WAL electron analyzer,
monochromatic laboratory sources and an open cycle helium cryostat.
The energy and angle resolution were set to 10 meV and 0.1$^o$ to
have adequate signal to noise ratio without compromising the
resolution necessary for this study. The sample was cleaved several
times \emph{in situ} at the experimental temperature using a top
post glued on top of the sample. The well ordered sample surface has
been verified by bright sharp low energy electron diffraction spots.
$X$-ray photoemission spectra from freshly cleaved sample was found
clean with no oxygen or carbon related signals.

The electronic band structure of a slab of Bi$_2$Se$_3$ was
calculated employing \emph{state of the art} full potential
linearized augmented plane wave method using Wien2k
software\cite{Wien2k}. The bulk electronic structure exhibits a gap
consistent with its insulating behavior \cite{dos_larson,dos_mishra}
and the results from slab calculations show signature of Dirac cone
in the presence of spin-orbit coupling.

\section*{Acknowledgements}

K. M. acknowledges financial support from the Dept. of Science and
Technology, Govt. of India under the `Swarnajayanti Fellowship
programme', the Dept. of Atomic Energy, Govt. of India, and the
Alexander von Humboldt Foundation, Germany. G. B. wishes to thank
EPSRC, UK for financial support through Grant EP/I007210/1.


\section*{AUTHOR CONTRIBUTIONS}
D. B. and S. T. carried out the photoemission measurements. D. B.
carried out band structure calculations and helped in data analysis.
K. A. helped in the experiments. G. B. prepared the sample and
characterized them. K. M. initiated the study, supervised the
measurements, analyzed the data, prepared the figures \& the
manuscript.

\section*{COMPETING FINANCIAL INTERESTS}

The authors declare no competing financial interests.

\end{document}